\documentclass[a4paper,12pt]{article}
\usepackage{epsfig}
\usepackage{graphicx}
\usepackage{amsmath}
\usepackage{amsfonts}
\usepackage{dsfont}
\usepackage{amssymb}
\usepackage{mathrsfs}
\usepackage{subfigure}
\usepackage{wasysym}
\newcommand{\be}{\begin{equation}}
\newcommand{\ee}{\end{equation}}
\newcommand{\bea}{\begin{eqnarray}}
\newcommand{\eea}{\end{eqnarray}}

\newcommand{\mbf}[1]{\mbox{\boldmath $#1$}}

\newcommand{\brho}{\mbf{\rho}}

\title{\textbf{Colored Spin Systems, BKP Evolution and finite $N_c$ effects
}
}
\author{ 
  \Large{P.~L.~Iafelice\footnotemark[1]~ and 
         G.~P.~Vacca\footnotemark[2]}\\[6pt]
  \textit{Dip.~di Fisica - Universit\`a di
             Bologna and INFN - Sezione di Bologna,}\\[0pt]
  \textit{via Irnerio 46, 40126 Bologna, Italy}}
\date{\today}
\begin{document}
\footnotetext[1]{email: iafelice@bo.infn.it}
\footnotetext[2]{email: vacca@bo.infn.it}
\maketitle
\begin{abstract}
\noindent
Even within the framework of the leading logarithmic approximation
the eigenvalues of the BKP kernel for states of more than three
reggeized gluons are unknown in general, contrary to the planar limit case
where the problem becomes integrable.
We consider a $4$-gluon kernel for a finite number of colors and define
some simple toy models for the configuration space dynamics, which
are directly solvable with group theoretical methods. Then we study the
dependence of the spectrum of these models with respect to the number of
colors and make comparisons with the large limit case. 
 \end{abstract}
\section{Introduction}
\label{intro}
In quantum field theory and statistical mechanics the $1/N$ (or large $N$)
expansion~\cite{tHooft} is a well known and extensively used perturbative
framework whenever
the theories under investigation present an internal symmetry
typically related to groups like $SO(N)$ or $SU(N)$.

Quantum Chromodynamics is one of the theories mostly studied under this
approximation even if, as a physical gauge theory, it is characterized by a gauge group
$SU(N_c)$ where the number of colors $N_c$ is just $3$.
Recently, thanks to the renewed interest induced by the ADS/CFT
correspondence, the $N=4$ SYM theory in the infinity color (planar) limit has
been intensively studied and several important results achieved.

The fact that the planar $N=4$ SYM is expected by the theoretical community to be
solvable and that it is dual to a superstring sigma model has led several
theorists to look for hints, in the absence of any supersymmetry, for the
existence of a possible dynamical system dual to planar QCD sharing with it
some integrability properties. The starting points are the integrable structures
unveiled many years ago at one loop in standard perturbation theory and
some hints of possible integrability at two loops in the planar limit.

The first evidence of an integrable structure at one loop in QCD was
 found~\cite{integrab} by L.N. Lipatov in the framework of the Regge limit
of scattering amplitudes whose behavior may be conveniently described by
systems of interacting reggeized gluons, as we shall briefly review in the next section.
The integrable dynamics, associated to the evolution in rapidity of such a
system, appears when one is taking  the large $N_c$ approximation, which makes
the BKP kernel~\cite{BKP} to resemble the structure of an Heisenberg XXX spin
 chain, but for a non compact $SL(2,{\mathbb C})$ ``spin''. 
 
Going beyond the large $N_c$ approximation, even in the lowest orders in
perturbation theory in the coupling constant, is a formidable task and it is
very difficult also to try to estimate the error one faces when computing
quantities for infinite $N_c$ (planar limit) instead of at $N_c=3$.

It is the purpose of this work to introduce some finite toy models, which
share the same color structure of the BKP systems and can be studied to determine
the dependence of the spectrum on the number of colors $N_c$. They are characterized by a
configuration space which is no more the transverse plane but a finite vector
space associated to irreducible representations of the $SU(2)$ group so that
one may use group theoretical methods to analyse some of these models.

This is of course not providing any concrete answer for the question related to the
real QCD problem, but nevertheless can be of some help. Moreover some toy models may be
interesting by themselves as dynamical systems.

We start in the next section with a short review of the properties of the
system of interacting reggeized gluons in the Leading Logarithmic
Approximation. In section three we consider the color structure for the four
reggeized gluon system and describe how to use a convenient basis for it.
In section four we construct some finite toy models which are studied in some
details in a couple of subsections. After the conclusions in few appendices we
give more details on symmetry structures and on the features of these toy models.
\section{BKP Kernels}
Let us start by giving a brief overview of the kernels which encode the
evolution in rapidity of systems of interacting reggeized gluons in the
leading logarithmic approximation (LLA). 
The reggeized gluons provide a convenient perturbative description of part of the
QCD degrees of freedom in the Regge limit (also known as the small $x$ limit)
and appeared in the investigations of the leading dependence of the
total cross sections on the center of mass energy in the LLA, which is
associated to the so called BFKL (perturbative) pomeron~\cite{BFKL}. 
In the simplest form, the BFKL pomeron turns out to be a composite state of
two interacting reggeized gluons ``living'' in the transverse configuration plane
in the colorless configuration.
The construction of the kernel reflects the property that in the Regge limit
the scattering amplitude factorizes in the impact factors which determine the
coupling of the external particles to the $t$-channel reggeized gluons and in
a Green's function which exponentiates the kernel and contains the rapidity
dependence of their composite state.
Such a dependence can be analyzed in terms of the spectral
properties of the kernel and in particular one is interested in the
eigenvalues and eigenstates associated to the leading behavior.
Because of this the spectral problem is often formulated in quantum mechanical
terms with the kernel being the ``Hamiltonian'' and its eigenvalues the
``energies''.

In the case of a colorless exchange the Hamiltonian is infrared finite and
in LLA is constructed summing the perturbative contributions of different
Feynman diagrams: in particular the virtual ones (reggeized gluon trajectories)
$\omega$ and the real ones (associated to an effective real gluon emission
vertex) $V$. One
writes formally $H=\omega_1+\omega_2+ \vec{T}_1 \vec{T}_2 V_{12}$
where $\vec{T}_i$ are the generators of the color group in adjoint
representation . In the colorless case one has $\vec{T}_1 \vec{T}_2 =-N_c$
and finally one obtains:
\be
H_{12}=\ln \,\left| p_{1}\right| ^{2}+\ln \,\left| p_{2}\right| ^{2}+\frac{1%
}{p_{1}p_{2}^{\ast }}\ln \,\left| \rho _{12}\right| ^{2}\,p_{1}p_{2}^{\ast
}\,+\frac{1}{p_{1}^{\ast }p_{2}}\ln \,\left| \rho _{12}\right|
^{2}\,p_{1}^{\ast }p_{2}-4\Psi (1)\,,
\label{BFKL_mom}
\ee
where $\Psi (x)=d\ln \Gamma (x)/dx$, a factor $\bar{\alpha}_s =\alpha_s
N_c/\pi$ has been omitted and the gluon
holomorphic momenta and coordinates have been introduced.

The gauge invariance gives the freedom
to choose a description within the M\"obius space~\cite{lipatov-moebius,BLV2},
wherein the functions describing the positions of the two reggeized gluons in
the transverse plane are zero in the coincidence limit. 
In this space the BFKL hamiltonian has the property of the holomorphic
separability ($H_{12}=h_{12}+\bar{h}_{12}$). Moreover
a remarkable property is its invariance under the
M\"obius group, whose generators for the holomorphic sector
in the M\"obius space for the principal series of unitary representations
are given by:
\be
M_{r}^{3}=\rho _{r}\partial _{r}\,,\,\,M_{r}^{+}=\partial
_{r}\,,\,\,M_{r}^{-}=-\rho^2_{r}\partial _{r}\, .
\label{moebius_gen}
\ee
The associated Casimir operator for two gluons is
\be
M^2 = |\vec{M}|^2=-\rho _{12}^2\,\partial _1\,\partial _2\,,
\label{moebius_cas}
\ee
where $\vec{M}=\sum_{r=1}^{2}\vec{M}_{r}$ and $\vec{M}_r%
\equiv(M_r^+,M_r^-,M_r^3)$.
Due to this symmetry the holomorphic and antiholomorphic parts 
of the Hamiltonian
can be written explicitly in terms of the Casimir
operator: indeed one has, after defining formally $J(J-1)=M^2$, 
\be
h_{12}=\psi(J)+\psi(1-J)-2\psi(1) \,.
\ee
Labelled by the conformal weights
$h=\frac{1+n}{2}+i\nu$ ,\,\,$\bar{h}=\frac{1-n}{2}+i\nu$, where
 $n$ is the conformal spin and $d=1-2i\nu $ is the anomalous dimension
of the operator $O_{h,\bar{h}}(\mbf{\rho }_{0})$ describing
the compound state \cite{Lcft},
the eigenstates and eigenvalues of the full hamiltonian in
eq. (\ref{BFKL_mom}), $H_{12} E_{h,\bar{h}} = 2 \chi_h E_{h,\bar{h}}$,
are respectively given by:
\be
E_{h,\bar{h}}(\brho_{10},\, \brho_{20}) \equiv \langle \rho | h \rangle
=\left(%
\frac{ \rho _{12}}{\rho _{10}\rho _{20}}\right) ^{h}\left( \frac{\rho
_{12}^{*}}{ \rho _{10}^{*}\rho _{20}^{*}}\right) ^{\bar{h}}\,,
\label{pomstatescoord}
\ee
and
\be
\chi_h \equiv \chi(\nu,n)= \psi\left(\frac{1+|n|}{2}+i\nu\right)
+\psi\left(\frac{1+|n|}{2}-i\nu \right) -2\psi(1) \,.
\label{eigenvalues}
\ee
The leading eigenvalue, at the point $n=\nu=0$, has a value
$\chi_{max}=4\ln 2\approx 2.77259$, responsible for the rise of the total cross
section as $s^{\bar{\alpha}_s \chi_{max}}$, which corresponds to a strong
violation of unitarity.

Let us now consider the evolution in rapidity of composite states of
more than $2$ reggeized gluons~\cite{BKP}.
The BKP Hamiltonian in LLA acting on a colorless state
can be written in terms of the BFKL pomeron Hamiltonian and
has the form (see \cite{integrab})
\be
H_n=-\frac{1}{N_c} \sum_{1\leq k < l\leq n} \vec{T}_{k} \vec{T}_{l}
H_{kl}\,.
\label{bkp_K}
\ee
This Hamiltonian is conformal invariant but cannot be solved in general.
Nevertheless the case of three reggeized gluons, where the color structure
trivially factorizes, is solvable~\cite{integrab}
and different families of solutions were found~\cite{JW,BLV}.
Physically these states are associated to the so called odderon
exchange~\cite{odderon} and in particular the family of solutions given in~\cite{BLV}
corresponds to eigenvalues up to zero (intercept up to one) and
are the leading one in the high energy limit.
Moreover they have a non null coupling to photon-meson impact
factors~\cite{BBCV}.

The case of more than three reggeized gluon is in general not solvable but if
one considers the color cylindrical topology when taking the large $N_c$ limit
the resulting Hamiltonian
\be
H_n^{\infty} = \frac{1}{2} \left[ H_{12}+ H_{23}
+ \cdots + H_{n1} \right] = h_n+\bar{h}_n
\label{bkpnlarge}
\ee
is integrable, i.e. there exists a set of other $n-1$ operators $q_r$,
which commute with it and are in involution. They are given,
in coordinate representation, by
\be
q_r=\sum_{i_1<i_2< \cdots < i_r} \rho_{i_1 i_2}\rho_{i_2 i_3} \cdots
\rho_{i_r i_1}\, p_{i_1}p_{i_2} \cdots p_{i_r} \,,
\ee
together with similar relations for the antiholomorphic sector. In particular,
$q_2=M^2$ is the Casimir of the M\"obius group.
This is the first case where integrability was found within the
context of gauge theories analyzing the Green's function in some
kinematical limit.
This integrable model is a non compact generalization of the Heisenberg XXX
spin chains~\cite{integrab,FK} and has been intensively studied with different
techniques in the last decade~\cite{dkm,dv-lip,dkkm,dv-lip2,vacca,kot}.

Here we remind the result for the highest eigenvalue of a system of four
reggeized gluons in the planar, one cylinder topology (1CT), case:
$H_4^{\infty} \psi_4=2 E_4^{\,1 \rm CT} \psi_4$.
The maximum value found, for zero conformal spin, is
\be
E_4^{\,1 \rm CT}=0.67416 \,.
\label{eigen1CT}
\ee
In general for an arbitrary number $n$ of reggeized gluon in the cylindrical
topology the leading eigenvalues have been found to be positive for even $n$ and
negative for odd $n$ and asymptotically behaving as $1/n$~\cite{dkkm,dv-lip2}.

The following are two important question that are unfortunately very hard to
answer: what are the
eigenvalues at finite $N_c=3$ and what is in general their dependence in
$N_c$? One may be tempted to apply variational or perturbative
techniques to the spectral problem, which nevertheless appears to be quite involved.
In any case a first step consists of analyzing the color structure, which
simplifies a bit in the case of four reggeized gluons in a total colorless
state.
\section{Color structure}
We consider the BKP kernel $H_4$ for four gluons, given in
eq. (\ref{bkp_K}). This is an operator acting on 4-gluon states, which may be
represented  as functions of the transverse plane coordinates and of the
gluon colors $v(\{\brho_i\})^{a_1 a_2 a_3 a_4}$. Let us concentrate here
on the color space.

It is convenient, due to the fact that the four gluons are in a total color
singlet state, to write the color vector $v^{a_1 a_2 a_3 a_4}$
in terms of the color state of a two gluon subchannel.
Let us therefore start from the resolution of unity for a state of two
$SU(N_c)$ particles in terms of the projectors $P[R_i]_{a_1 a_2}^{a'_1 a'_2}$
onto irreducible representations: 
\be
1=P_1+P_{8A}+P_{8S}+P_{10+\bar{10}}+P_{27}+P_{0} = \sum_i P[R_i]\,,
\ee
where $Tr P[R_i]=d_i$ is the dimension of the corresponding representation.
Let us note that we have chosen to consider a unique subspace for the direct sum
of the two spaces corresponding to $10$ and $\bar{10}$ representations. This
is convenient for our purposes and we shall therefore consider just $6$
different projectors to span the color space of two gluons.  

If we consider gluons $(1,2)$ to be the reference channel we introduce as the
base for the color vector space the set $\{ P[R_i]_{a_1 a_2}^{a_3 a_4}\}$ of
projectors and write
\be
v^{a_1 a_2 a_3 a_4}=\sum_i v^i \left(P[R_i]_{a_1 a_2}^{a_3 a_4}\right)  \quad
\mathrm{or} \quad v=\sum_i v^i P_{12}[R_i]\,.
\ee
Note that one could have also chosen other reference channels corresponding to
a description in terms of projection onto irreducible representations of other
gluon subsystems.
Having chosen a color basis, we find that the next step is to write the BKP kernel with
respect to it. We can slightly simplify the expression for the kernel
since for a colorless state we have $\sum_i \vec{T}_i v=0$ which
implies that $\vec{T}_1\vec{T}_2 v= \vec{T}_3\vec{T}_4 v$ (an similarly for
the other permutations of the indices). Therefore one may write:
\be
H_4=-\frac{1}{N_c} \left[ \vec{T}_1\vec{T}_2
  \left(H_{12}+H_{34}\right)+\vec{T}_1\vec{T}_3 \left(H_{13}+H_{24}\right)+
\vec{T}_1\vec{T}_4 \left(H_{14}+H_{23}\right) \right].
\label{bkp_K_2}
\ee
Let us now write explicitly the action of the color operators $\vec{T}_i\vec{T}_j=\sum_a
T^a_i T^a_j$ which are associated to the interaction between the gluons
labelled $i$ and $j$.
We start from the simple ``diagonal channel'' for which we have relation
$\vec{T}_i\vec{T}_j= -\sum_k a_k P_{ij}[R_k]$ with coefficients
$a_k=(N_c,\frac{N_c}{2},\frac{N_c}{2},0,-1,1)$.
Consequently we can write in the $(1,2)$ reference base
\be
\left(\vec{T}_1\vec{T}_2 v \right)^j=- a_j v^j = -\left( A \,v \right)^j \,,
\label{action12}
\ee
where $A=diag(a_k)$.
The action on $v$ of the $\vec{T}_1\vec{T}_3$ and $\vec{T}_1\vec{T}_4$
operators is less trivial and is constructed in terms of the $6j$ symbols of the
adjoint representation of $SU(N_c)$ group. We shall give few details in the
appendix A and write directly the results, in terms of the symmetric (after a
similarity transformation) matrix operators:
\be
\left( \vec{T}_1\vec{T}_3 \,v \right)^j=- \sum_i \left( \sum _k C^j_k a_k
    C^k_i\right) v^i = -\left( {C A\, C}\, v\right)^j
\label{action13}
\ee
and
\be
\left( \vec{T}_1\vec{T}_4 \,v \right)^j=- \sum_i \left( \sum _k s_j C^j_k a_k
    C^k_i s_i \right) v^i = -\left( {S C A\, C S}\, v\right)^j \,.
\label{action14}
\ee
The matrix $C$ is the crossing matrix built on the $6j$ symbols and
$S=diag(s_j )$ is constructed on the parities $s_j=\pm 1$ of the different
representations $R_j$.

We can therefore write the general BKP kernel for a four gluon state,
given in eq. (\ref{bkp_K_2}), as
\be
H_4=\frac{1}{N_c} \left[ A \left(H_{12}+H_{34}\right)+C A
  C\left(H_{13}+H_{24}\right)+
S C A C S \left(H_{14}+H_{23}\right) \right]
\label{H4colordecomp}
\ee
One can check that if we make trivial the transverse space dynamics, replacing
the $H_{ij}$ operators by a unit operators, the general BKP kernel in
eq. (\ref{bkp_K}) becomes $H_n= \frac{n}{2} \hat{1}$ and indeed one can verify
that $A+C A C+ S C A C S=N_c \hat{1}$.

Let us make few considerations on the large $N_c$ limit approximation. As we
have already discussed, in the Regge limit one faces the factorization of an
amplitude in impact factors and a Green's function which exponentiates the
kernel.
The topologies resulting from the large $N_c$ limit depend on the impact factor
structure. In particular one expects the realization of two cases: the one
and two cylinder topologies. The former corresponds to the case, well studied, of the
integrable kernel, Heisenberg XXX spin chain-like. It is encoded in the relation:
$\vec{T}_i \vec{T}_j\to -\frac{N_c}{2}\delta_{i+1,j}$ which leads to
 $H_4=\frac{1}{2}\left(H_{12}+H_{23}+H_{34}+H_{41}\right)$.
It is characterized by eigenvalues corresponding to an intercept less then a pomeron.
The latter case instead is expected to have a leading intercept, corresponding
to an energy dependence given by two pomeron exchange. Consequently one
expects at finite $N_c$ a contribution with an energy dependence even stronger. 
In the two cylinder topology the color structure is associated to two singlets
($\delta_{a_1a_2}\delta_{a_3a_4}$, together with
the other two possible permutations). Such a structure is indeed present in
the analysis, within the framework of extended generalized LLA, of unitarity
corrections to the BFKL pomeron exchange~\cite{Bartels} and diffractive
dissociation in DIS~\cite{BW}, where the perturbative triple pomeron
vertex (see also~\cite{BV}) was discovered and shown to couple exactly to the
four gluon BKP kernel. 

It is therefore of great importance to understand how much the picture derived
in the planar $N_c=\infty$ case is far from the real situation with
$N_c=3$. One clearly expects for example that the first corrections to the
eigenvalues of the BKP kernel are proportional to $1/N_c^2$, but what is
unknown is the multiplicative coefficient as well as the higher order terms. 
\section{Toy models}
In this section we shall consider a family of models, different from
the BKP system, which neverthelss share several features with it and can
be used to judge how the large $N_c$ approximation might be more or less
satisfactory.
Moreover these systems may be considered interesting by themselves as quantum
dynamical systems.

A state of $n$ reggeized gluons undergoing the BKP evolution, described by the
kernel in eq. (\ref{bkp_K}), belongs to a vector space of functions on a domain
given by the tensor product of the color space $\bf 8^n$ and
the configuration space $\mathbb R^{2n}$, associated to the position or
momenta in the transverse plane, of the $n$ gluons.
Indeed the BKP kernel is built as a sum of product of
color ($ \vec{T}_{k} \vec{T}_{l}$) and of configuration ($H_{kl}$) operators;
the latter, on the M\"obius space, can be written in terms of the
Casimir of the M\"obius group, i.e. in terms of the scalar product of the
generators of the non compact spin group $SL(2,{\mathbb C})$: $H_{kl}=H_{kl}(
\vec{M}_{k}\cdot \vec{M}_{l})$. 

We are, therefore, led to consider a class of toy models where the BKP configuration
space $\mathbb R^{2 n}$ is substituted by the space $V_s^n$ where $V_s$ is the
finite space spanned by spin states belonging to the irreducible
representation of $SU(2)$ with spin $s$. In particular we shall consider quantum
systems with an Hamiltonian fitting the following structure:
\be
{\cal H}_n=-\frac{1}{N_c} \sum_{1\leq k < l\leq n} \vec{T}_{k} \vec{T}_{l}\,
f(\vec S_k \vec S_l)\,,
\label{bkp_toy_K}
\ee
where $\vec S_i$ are the elements of the $su(2)$ algebra associated to the particle $i$ in
any chosen representation and $f$ is a generic function. A particular toy
model is therefore specified by giving the spin $s$ of each particle
(``gluons'') and the function $f$.
In the following we shall consider two specific cases for the $4$ particle system: 

a) A spin $s=1$ case in a global singlet state $v$ ($\sum_i \vec{S}_i \, v=0$). If $f$
is the identity map than the ``spin'' configuration dynamics is very similar
to the one of the color sector. In order to have a system which behaves similarly to the
BKP case we first put a constraint on the two particle operators, which
describe the basic interaction.
In particular we consider the family of functions
\be
f_\alpha(x)=
2{\rm Re}\left[ \psi \left(\frac{1}{2}+\sqrt{-\alpha (4+2 x)}\right)\right]
-2\psi(1) \,.
\label{f_form}
\ee
Remembering that for conformal spin $n=0$ the BFKL
Hamiltonian is given by
$H_{kl}=2{\rm Re}\left[ \psi \left(\frac{1}{2}+\sqrt{\frac{1}{4}
+(\vec{M}_{k}+\vec{M}_{l})^2}\right)\right]-2\psi(1)$,
one immediately recognizes that the $f_\alpha$ is associated to the
substitution $\frac{1}{4}+L^2_{ij} \to - \alpha\, S_{ij}^2$ which assures to have the
same leading eigenvalue for any $\alpha$, since both expressions have the
value zero as upper bound. The parameter $\alpha$ will be
chosen in order to constrain the full 4-particle Hamiltonian (\ref{bkp_toy_K})
to have the same leading eigenvalue as the QCD BKP system in the large $N_c$
limit (at zero conformal spin).
In this system, the BKP toy model, we shall investigate finite $N_c$ effects. 

b) A system ${\rm TOY_{Adj,Fund}}$
with $f$ the identity function and spin $s=1/2$. Such a system in
the large $N_c$ limit in the case of one cylinder topology becomes the well
known Heisenberg XXX spin chain system which is integrable. We shall perform
some check on the $N_c$ dependence again for the $4$-particle case. 

c)
Moreover in order to have another check of the approach we shall also consider a model
where the $4$ particle belong to the fundamental representation of $SU(2)$ for
both the ``color'' and the ``spin'' so that we can perform a comparison with
standard results from the spectroscopy of isospin-spin systems.  
We place these checks in the appendix C. 
\subsection{BKP toy model}
In order to explicitly study this finite system, described by the Hamiltonian
in eq. (\ref{bkp_toy_K}) acting on vector states with dimension $(8 \times
3)^4$ and singlet under both $SU(3)_C$ and $SU(2)_{spin\,\, conf}$,
it is convenient to choose the color decomposition in 2-particle subchannel irreducible
representations described in section $3$ and adopt a similar
approach also on the ``spin'' degrees of freedom. 
After that one is left with the problem of diagonalizing an Hamiltonian which
is a matrix $18 \times 18$, a problem addressable with any computer.
Without the singlet restriction on the spin part the problem in general is
much more complicated to be easily solved and may be addressed in future investigations.

Let us therefore proceed by introducing for 2 particle spin $1$ states the
resolution of unity $1=Q_1+Q_3+Q_5=\sum_i Q[R_i]$ which let us write
$\vec{S}_i \vec{S}_j=-\sum_k b_k Q_{ij}[R_k]$ with $b_k=(2,1,-1)$ (c.f. with
$a_k$: first, second and second last terms).
It is, therefore, straightforward to write from a power series representation
($Q_{ij}[R_k]$ are projectors):
\be
f(\vec{S}_i \vec{S}_j)=\sum_k f(-b_k) Q_{ij}[R_k]\,.
\ee
Using the corresponding crossing matrices $D$ and the parity matrix $S'$
one obtains relations very similar to the one reported in
eqs. (\ref{action12})-(\ref{action14}), which read
\be
\left(f\left(\vec{S}_1\vec{S}_2 \right)v \right)^j=f(- b_j) v^j =\left( B \,v \right)^j \,,
\label{action12p}
\ee
\be
\left( f\left(\vec{T}_1\vec{T}_3\right) \,v \right)^j= \sum_i \left( \sum _k D^j_k f(-b_k)
    D^k_i\right) v^i = \left( {D B\, D}\, v\right)^j
\label{action13p}
\ee
and
\be
\left(f\left( \vec{T}_1\vec{T}_4 \right)\,v \right)^j=
\sum_i \left( \sum _k s'_j D^j_k f(-b_k)
    D^k_i s'_i \right) v^i = \left( {S' D B\, D S'}\, v\right)^j \,.
\label{action14p}
\ee
From the above results for the two particle representation basis, we can
write the explicit form of the Hamiltonian for this toy model,
going beyond the one given in eq. (\ref{H4colordecomp}). Indeed we obtain
\be
{\cal H}_{4a}=\frac{2}{N_c}\left( A\otimes B+C A\, C \otimes D B\, D+   S C A\, C S
\otimes  S' D B\, D S' \right)
\label{H4BKPtoy}
\ee
which contains a dependence on $N_c$ and on the parameter $\alpha$ through the
function $f_\alpha$ given in eq. (\ref{f_form}).

Let us note that in the large $N_c$ limit one faces for the Hamiltonian two
possible cases: the one cylinder topology (1CT) which corresponds to the
simpler Hamiltonian
\be
{\cal H}_{4a}^{1CT}=-\frac{1}{N_c}\left[ -\frac{N_c}{2} \sum_i f\left(\vec{S}_i
  \vec{S}_{i+1} \right)\right]= B+S'DBDS' 
\label{H4a1cyl}
\ee
and the two cylinder topology (2CT) corresponding to the even simpler
  Hamiltonian
 \be
{\cal H}_{4a}^{2CT}=-\frac{1}{N_c}\left[ -N_c f\left(\vec{S}_1
  \vec{S}_2 \right)-N_c  f\left(\vec{S}_3 \vec{S}_4 \right)\right]=2B \,.
\label{H4a2cyl}
\ee
Let us remark that while in the case of $N_c>3$ we consider a basis for the
vector states made of
{$P[R_i]Q[R_j]$} with $18$ elements since in the color sector there is also
the $P_0$ projector, the case $N_c=3$ is characterized by a basis of $15$
elements.

As already anticipated, in order to study a toy model resembling the spectrum
of the BKP system of $4$ gluons, we require that, in the large $N_c$ limit
in the one cylinder topology, the leading eigenvalue must be the same as
the one found for the corresponding integrable BKP system, whose value was given in
eq. (\ref{eigen1CT}). This fact fixes the value of the parameter $\alpha=2.80665$.  
We are therefore left with an Hamiltonian which is just a function of the
number of colors $N_c$. 

Let us now consider its spectrum for the cases $N_c=3$
and $N_c=\infty$. Here we report the values followed by their
multiplicities. Note than for $N_c=3$ there are $15$ eigenvalues while they are
$18$ for any other value of $N_c$. For the case $N_C=\infty$ we
specify also the topology they belong to.

\begin{center}
\begin{minipage}{5cm}
\[\left(\begin{array}{c}
N_c=3\\
{\bf 7.04193}\quad (\times 1)\\ {\bf 5.51899} \quad (\times 2)\\ 
{\underline{1.12269}}\quad (\times 2)\\ 
-3.89328\quad (\times 2)\\-4.04744\quad (\times 1)\\
-4.27838\quad (\times 1)\\ -7.81242\quad (\times 1)\\ -9.18576\quad
(\times2)\\ -12.6743\quad (\times 2)\\ -14.1005\quad (\times 1)
\end{array}
\right)
\]
\end{minipage}
{$\to$}
\begin{minipage}{5cm}
\[\left(\begin{array}{c}
N_c=\infty\\
{\bf 5.54518}\quad (\times 3) \,\, 2CT \\ {\underline{ 0.67416}}\quad (\times 3) \,\,1CT\\
-4.27838\quad (\times 3) \,\,1CT\\ -7.81242\quad (\times 3) \,\,2CT\\
-8.67983\quad (\times 3) \,\,1CT\\ -10.0168\quad (\times 3) \,\,2CT\\
\end{array}
\right)
\]
\end{minipage}
\end{center}

\vspace*{0.3cm}
We track the flow from $N_C=3$ to $N_c=\infty$: the first three
highest eigenvalues (in bold) are moving to the same leading value (in
bold) which corresponds to two BFKL pomeron exchange (in two
cylinder topology). The fourth and fifth highest eigenvalues (underlined) are instead
moving to the leading eigenvalues of the one cylinder topology case (which are
three instead of two because of the larger basis for $N_c>3$). 
With very good approximation one finds that the $N_c$ dependence of the
leading eigenvalue $E_0$ is given by
\be
E_0(N_c)=E_0(\infty)\left(1+\frac{2.465}{N_c^2}\right) \,.
\ee
One can see that for this toy model the large $N_c$
approximation corresponds to an error of about $27\%$, an error which is not negligible
because the coefficient of the leading correction to the asymptotic value, proportional to
$1/N_c^2$, is a large number.

It is also easy to investigate the color-configuration space mixing which is
encoded in the eigenvectors. We report some results in the appendix B.
\subsection{${\rm TOY_{Adj,Fund}}$}
We now move to study the toy model described at point (b) at the beginning of
section $4$, again to see how the large $N_c$ approximation works.
It is described by the Hamiltonian
\be
{\cal H}_{Adj,Fund}=-\frac{1}{N_c} \sum_{1\leq k < l\leq n} \vec{T}_{k} \vec{T}_{l}
\frac{\vec \sigma_k}{2} \frac{\vec \sigma_l}{2}\,,
\label{toy_b}
\ee
acting on spin singlet states. 
Again we consider the large $N_c$ limit. The one cylinder topology is
associated to the well known Heisenberg XXX spin chain with Hamiltonian
\be
{\cal H}_{Adj,Fund}^{1CT}= \frac{1}{2} \sum_{i=1}^n 
\frac{\vec \sigma_i}{2} \frac{\vec \sigma_{i+1}}{2}\,,
\label{toy_b1CT}
\ee
which we shall now consider for the case of $n=4$ particle. In this case at
large $N_c$ we have, as before, also the two cylinder topology associated to the Hamiltonian
\be
{\cal H}_{Adj,Fund}^{2CT}= 2  
\frac{\vec \sigma_1}{2} \frac{\vec \sigma_{2}}{2}\,.
\label{toy_b2CT}
\ee
The spectrum for the one cylinder topology case is well known from Bethe
Ansatz methods~\cite{Faddeev} and for total zero spin of a $4$-particle spin
chain the possible eigenvalues are $0$ and $-1$ (see table II
in~\cite{Karbach} for $J=-1/2$ in their notation). The two cylinder topology
is characterized by the eigenvalues $+1/2$ and $-3/2$.

\begin{figure}
\begin{center}
\resizebox{.6\textwidth}{!}{\includegraphics{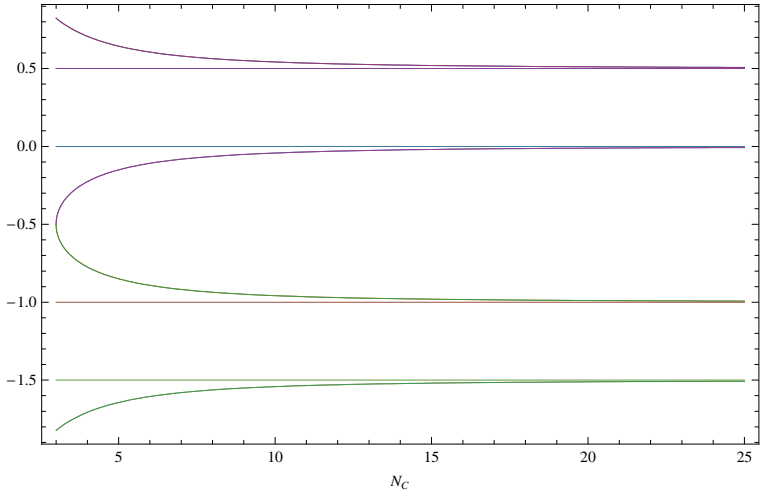}}
\end{center}
\caption{$N_c$ dependence of the eigenvalues of the model ${\rm
    TOY_{Adj,Fund}}$.}
\label{fig1}
\end{figure}

At finite $N_c$ we rewrite the Hamiltonian in a similar way to the BKP toy
model case (see eq. (\ref{H4BKPtoy}) where the $B$ and $D$ matrices are defined for $f$
the identity map and for the group $SU(2)$ in fundamental representation).  
At $N_c=3$ it corresponds to a $10\times 10$ matrix while for $N_c>3$ it is
given by a $12\times12$ matrix.
The leading eigenvalue as function of $N_c$ can be easily computed
\be
E_0(N_c)=\frac{\sqrt{10 N_c^2 +36+6\sqrt{N_c^4+36N_c^2+36}}-2N_c}{4N_c}
\ee
and indeed goes to the value $1/2$ in the large $N_c$ limit.
Let us note that if one considers the planar approximation (in the 2CT configuration),
the leading eigenvalue would be underestimated with a relative error of
$\left(E_0(3)-E_0(\infty)\right)/E_0(3)\simeq 40\%$ w.r.t. the case $N_c=3$.
In Fig. \ref{fig1} we
report the $N_c$ dependence of all the eigenvalues in the range $3\le N_c \le 25$.
 
Similar models, but in a higher spin representation, can be constructed in
order to maintain the integrability in the large $N_c$ limit. One simply needs
to consider for any irreducible representation $s$ of the particles the
function $f$ to be a corresponding specific polynomial as described in~\cite{Faddeev}.
\section{Conclusions}
We have introduced a family of dynamical models describing interacting particles with
color and spin degrees of freedom. The main motivation was to study within this
framework how much the large $N_c$ approximation is significant when one is trying to
extract the spectrum of these quantum systems.

Indeed in some relevant physical cases the only results available are
restricted to the case with a planar structure resulting from the
large $N_c$ approximation, when integrability arises and
gives the possibility to exactly solve the problem. These facts are seen
when considering QCD scattering amplitudes in the Regge limit and LLA
approximation, characterized by the BKP dynamics.

We have focused our study to the the case of four particles and considered in
details three toy models.
One toy model (case (c) in section $4.2$) was considered to test our
computational method based on group theory since one is able to make a direct comparison
with results already known from other methods used in spectroscopy.

The first model presented in section $4.2$ (a) is aimed to mimic to some extend the
behavior of the $4$ gluon BKP kernel,  since we have forced it to have in the
large $N_c$ limit the same leading eigenvalues of the BKP system for both one
and two cylinder topologies.
We were able to compute the different eigenvalues of this toy
model as function of $N_c$ and we have found that the leading one at $N_c=3$
present corrections of almost $30\%$ w.r.t. the planar approximation, which
one may understand in terms of a large coefficient in the $1/N_c^2$ correction term. The
mixing in color-spin configuration structure has been also studied.

Another model (case (b) in section $4.2$) was considered since in the large
$N_c$ limit it gives rise to the one cylinder topology Heisenberg XXX spin
chain which is integrable. For the spin $1/2$ case we have found at finite
$N_c=3$ corrections to the leading eigenvalue of about $40\%$.

Let us note that in our analysis we were restricting ourselves to study the
toy model Hamiltonians on the space of states which are singlet with respect to the
$SU(2)$ ``spin'' configurations. This was a choice dictated by technical
reasons but one should look forward to extending the investigation to all the
possible states.

These kind of models and possibly more general ones appear to be interesting
also by themselves and we feel that they deserve more studies in order to see,
for example, if some remnant from integrability can be traced back at finite $N_c$.
\section*{Acknowledgements}
G.P.V. thanks the Alexander von Humboldt Stiftung for the support during the
early stage of this work and is very grateful to J. Bartels, L.N. Lipatov
M. A. Braun and M. Salvadore for very interesting and stimulating discussions.
\section*{Appendix A}
In this appendix we note a few facts about the crossing matrices introduced in
section $3$ and $4$ for the $SU(N_c)$ group. Related considerations may be found in
\cite{RS,cvit,NSZ,DM,KL} where explicit expressions for the crossing matrices
can be found and therefore will not be given here.

Let us rewrite in graphic notation the operator $\vec{T}_i\vec{T}_j$ in the
basis $(i,j)$ and  the color vector state in the basis $(1,2)$.

\resizebox{.45\textwidth}{!}{\includegraphics{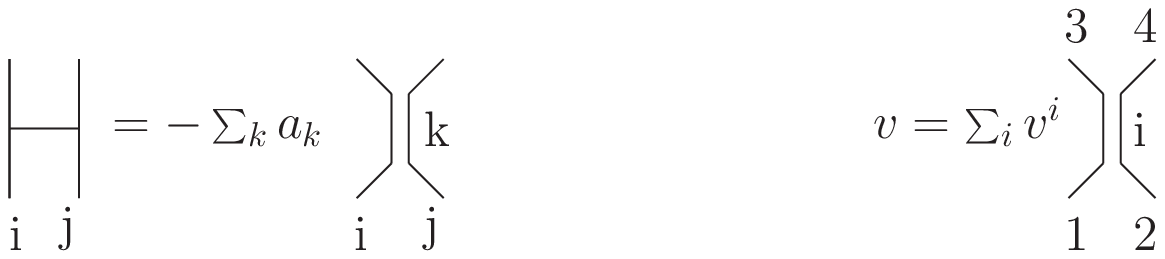}}

Let us compute the first non trivial crossing case, $\vec{T}_1\vec{T}_3 \, v$,
remembering to rewrite the final result again in the basis $(1,2)$. In a
graphical notation we have

\resizebox{.8\textwidth}{!}{\includegraphics{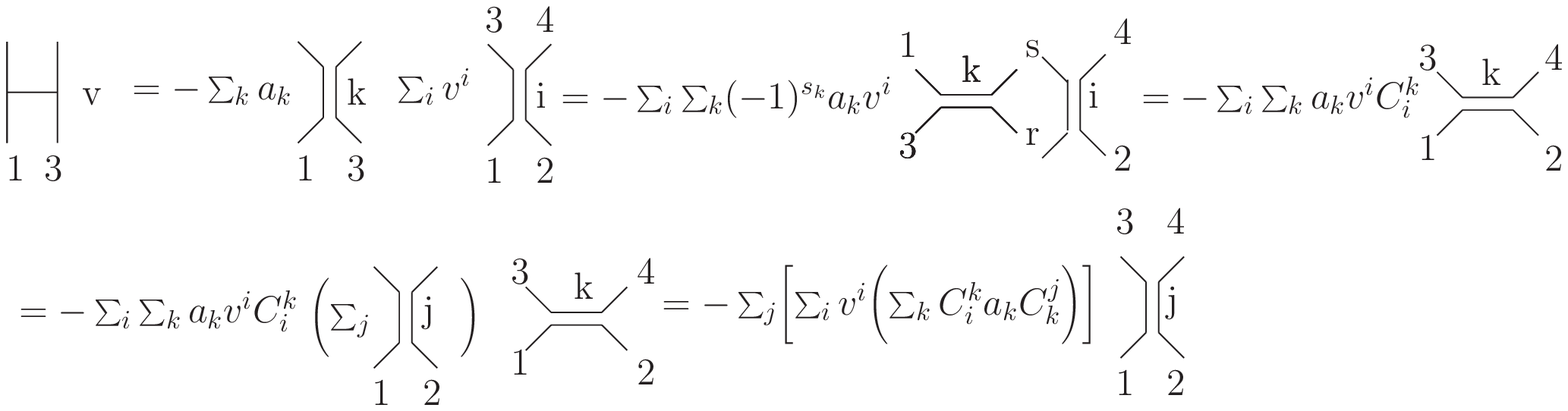}}

where the crossing matrix (essentially 6j symbols) can be written as

\resizebox{.2\textwidth}{!}{\includegraphics{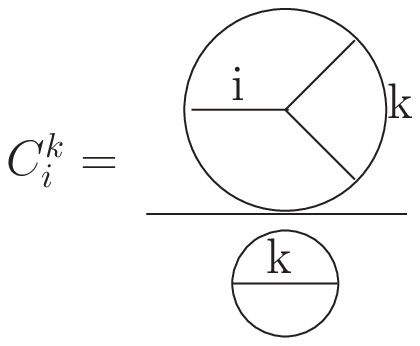}}

In a similar way one can also trace the action of the $\vec{T}_1\vec{T}_4$ operator.
One can see that in the last relation there is an asymmetry due to the fact
that one divides by the dimension of the $k$-representation. 
It is convenient to perform a similarity transformation to work with a
symmetric crossing matrix. For this purpose
it is sufficient to introduce the matrix $\Delta=\,$diag$(d_i)$ and define the new
symmetric matrix $C \to \Delta^{-\frac{1}{2}} C \Delta^{\frac{1}{2}}$ which
acts on the vectors with components  $v^i \to \left(\Delta^{-\frac{1}{2}}
  v\right)^i$ . 
\section*{Appendix B}
Let us consider the BKP toy model described and analysed in section $4.1$.
From numerical investigations one finds that the leading eigenvector $v_0$ and
the two closest subleading $v_{1,2}$ at $N_c=3$ have the following components 

\noindent
\begin{minipage}{5cm}
\[
v_0\simeq\left( \begin{array}{cc}
0.590 & P_1 Q_1\\ 0.085 & P_1 Q_5\\ 0.344 & P_{8A} Q_3\\
0.199 & P_{8S} Q_1\\0.199 & P_{8S} Q_5\\
0.293 & P_{10+\bar{10}} Q_3\\
0.179 & P_{27} Q_1\\0.574 & P_{27} Q_5
\end{array}
\right)
\]
\end{minipage}
\begin{minipage}{5cm}
\[
v_1 \simeq\left( \; \begin{array}{cc}
0.166 & P_1 Q_3\\0.342 & P_{8A} Q_1\\ 0.317 & P_{8A} Q_5\\
0.385 & P_{8S} Q_3\\0.267 & P_{10+\bar{10}} Q_1\\
0.598 & P_{10+\bar{10}} Q_5\\
0.421 & P_{27} Q_3
\end{array}
\right)
\]
\end{minipage}
\begin{minipage}{5cm}
\[
v_2 \simeq\left( \begin{array}{cc}
-0.775 & P_1 Q_1\\ 0.002 & P_1 Q_5\\ 0.008 & P_{8A} Q_3\\
0.123 & P_{8S} Q_1\\0.114 & P_{8S} Q_5\\
0.268 & P_{10+\bar{10}} Q_3\\
0.151 & P_{27} Q_1\\0.525 & P_{27} Q_5
\end{array}
\right)
\]
\end{minipage}

\vspace*{0.3cm}
As one can see the eigenvector $v_0$ of the highest eigenvalue is even which the
two fold degerate next lager eigenvalue has eigenstates of both parities
($v_1$ odd and $v_2$ even).

In the large $N_c$ limit case the eigenvectors of the three fold degenerate
leading eigenvalue of the 2 cylinder topology are

\noindent
\begin{minipage}{5cm}
\[
w^{(2CT)}_0\simeq\left( \begin{array}{cc}
1 & P_1 Q_1
\end{array}
\right)
\]
\end{minipage}
\begin{minipage}{5cm}
\[
w^{(2CT)}_1 \simeq\left( \; \begin{array}{cc}
\frac{1}{3} & P_{10+\bar{10}} Q_1\\
\frac{\sqrt{5}}{3} & P_{10+\bar{10}} Q_5\\
\frac{1}{\sqrt{6}} & P_{27} Q_3\\
\frac{1}{\sqrt{6}} & P_{0} Q_3
\end{array}
\right)
\]
\end{minipage} \ \ \
\begin{minipage}{5cm}
\[
w^{(2CT)}_2 \simeq\left( \begin{array}{cc}
\frac{1}{\sqrt{3}} & P_{10+\bar{10}} Q_3\\
\frac{1}{3 \sqrt{2}} & P_{27} Q_1\\
\frac{\sqrt{5}}{3 \sqrt{2}} & P_{27} Q_5\\
\frac{1}{3 \sqrt{2}} & P_{0} Q_1\\
\frac{\sqrt{5}}{3 \sqrt{2}} & P_{0} Q_5
\end{array}
\right)
\]
\end{minipage}

\vspace*{0.3cm}
Again also in this system we can track the same parity properties, which are
invariant under the flow in $N_c$.

Similarly one may investigate the states associated to the one cylinder
topology at $N_c=\infty$ and their corresponding partners at finite $N_c$.
For brevity we just report here the two most relevant states in the $N_c$
infinity limit: 

\noindent
\begin{minipage}{5cm}
\[
w^{(1CT)}_0\simeq\left(\!\! \begin{array}{cc}
z_1 & P_{8A} Q_1 \\
z_3 & P_{8S} Q_3 \\
z_5 & P_{8A} Q_5 
\end{array}
\!\right)
\]
\end{minipage}\!\!
\begin{minipage}{5cm}
\[
w^{(1CT)}_1 \simeq\left( \!\! \begin{array}{cc}
z_1 & P_{8S} Q_1 \\
z_3 & P_{8A} Q_3 \\
z_5 & P_{8S} Q_5 
\end{array}
\! \right)
\]
\end{minipage}\!
\begin{minipage}{5cm}
\[
w^{(1CT)}_2 \simeq\left( \!\! \begin{array}{cc}
0.245 & (P_0 Q_1-P_{27} Q_1) \\
0.663 & (P_0 Q_5-P_{27} Q_5)
\end{array}
\! \right)
\]
\end{minipage}

\vspace*{0.3cm} \noindent
where $z_1\simeq 0.815$, $z_3\simeq 0.405$ and $z_5\simeq 0.415$. We stress that
$w^{(1CT)}_2$ has no corrispective at $N_c=3$.
\section*{Appendix C}
This appendix is devoted to check in one specific case that our approach gives result
in agreement with other methods widely used in spectroscopy.
We start by considering a system of $n$ particles in the bifundamental representation of
$SU(N_c)\times SU(2)$, characterized by an Hamiltonian \eqref{bkp_toy_K} (with
$f$ the identity function)
\be
{\cal H}_n=-\frac{1}{N_c} \sum_{1\leq k < l\leq n} \vec{T}_{k} \vec{T}_{l}\;\;
\vec S_k \vec S_l\,,
\ee
 which can be written in terms of the quadratic
Casimir operators of $SU(N_c)$, $SU(2)$ and $SU(2N_c) \supset SU(N_c) \times SU(2)$ (see
\cite{Jaffe}).

Indeed the tensor products of $\vec{T}_{k} \vec S_l$ are among the
generators of $SU(2N_c)$, so it is useful to introduce the entire algebras for this
group
\be
\alpha_k=
\begin{cases}
\frac{1}{\sqrt{N_c}} S_l    & k=1,2,3=l\\
\frac{1}{\sqrt{2s+1}}T_{a}  & k=4,\dots,N_c^2+2;\; a=1,\dots,N_c^2-1\\
\sqrt{2}\, T_{k}  S_l       & k=N^2+3,\dots,4N_c^2-1; \; l=1,2,3
\end{cases}
\label{su4algebras}
\ee
with the normalization $Tr(\alpha_k\alpha_{k'})=1/2\,\delta_{kk'}$. 
The Hamiltonian for this system can be rewritten as
\be
{\cal H}_{All-fund}=-\frac{1}{4N_c}\left[\mathcal{C}_{2N_c}
  -\frac{1}{N_c}\mathcal{C}_{N_c} - \frac{1}{2s+1}\mathcal{C}_2 -2 n
  \,\frac{N_c^2-1}{2N_c} s(s+1)\right] \,,
\label{ham-toycasimir}
\ee
where the quadratic Casimir operators ${\cal C}_n$ are defined as
in~\cite{Jaffe} and $s=1/2$. Note that all the operators introduced above depend on the
irreducible representation of the symmetry group to which they refer to.

We are interested in the real representations so we set $N_c=2$ and
consider the case of only four particles. The symmetry
group of the model becomes $SU(2)\otimes SU(2) \subset SU(4)$ and
eq.~\eqref{ham-toycasimir}, written for the four particle in a global singlet state,
takes the form
\be
{\cal
  H}_{All-fund}=-\frac{1}{8}\left[\mathcal{C}_{4}(\mathcal{R})-\frac{9}{2}\right] \,.
\label{ham-toycasimirN=2}
\ee
In order to find its spectrum the next step consists of analyzing the
irreducible representation content of each symmetry group of the model.
So, for four particle with spin $1/2$, one has (we specify also the multiplicity)
\be
\bf2\otimes 2\otimes 2\otimes 2= \text{2}(1)+\text{3}(3)+5,
\ee
and in the $SU(4)$ case
\be
\bf4\otimes 4\otimes 4\otimes 4= 1+\text{3}(15)+\text{2}(20)+35+\text{3}(45).
\label{SU(4)decomp}
\ee
Then we need to study the $SU(2)\otimes SU(2)$ content of these $SU(4)$
$irrep$. This can be done using the results of \cite{itzykson} and in
particular the entries of table $1$,
\begin{table}[t]
\centering
 \begin{tabular}{|c||c||c|}
 \hline 
  $SU(2)\otimes SU(2)$ & $SU(4)$ & $(\mu_1,\mu_2,\mu_3)\equiv\mathcal{R}_SU(4)$ \\
 \hline \hline
  $\bf1\otimes 1$  & $1, 20, 35$      & $(0,0,0) (0,2,0) (4,0,0)$ \\ 
 \hline
  $\bf1\otimes 3$  & $15, 45$         & $(1,0,1) (2,1,0)$ \\
 \hline
  $\bf1\otimes 5$  & $20$             & $(0,2,0)$ \\
 \hline
  $\bf3\otimes 3$  & $15, 20, 35, 45$ & $(1,0,1)(0,2,0)(4,0,0)(2,1,0)$ \\
 \hline
  $\bf3\otimes 5$  & $45$             & $(2,1,0)$ \\
 \hline
 \end{tabular}
\label{irrep-decomposition}
\caption{Correspondence between {\it irreps} of $SU(4)$ and $SU(2) \otimes SU(2)$}
\end{table}
where the values in the third column are Dynkin indices.

So for particles in a total singlet state ($\bf1\otimes1$) the Hamiltonian
in eq.~\eqref{ham-toycasimirN=2} admits four eigenvalues, each for a different
$irrep$ of $SU(4)$, with a 2-fold degenerate eigenvalue corresponded to
$\bf20_{SU(4))}$ (see eq.~\eqref{SU(4)decomp}):
\be
\begin{cases}
  -\frac{15}{16},        & \text{for} \;irrep\;\bf 35\\
  -\frac{3}{16} \,\,\,\,(2\text{x}),    & \text{for} \;irrep\;\bf 20\\
  +\frac{9}{16} ,        & \text{for} \;irrep\;\bf 1
\end{cases}
\label{}
\ee
and these are in perfect agreement with the spectrum evaluated with the method
used previously throughout the paper (we take advantage from the formulas
of \cite{greiter} for the eigenvalues of a quadratic Casimir operator as
functions of the Dynkin indices). 

As a final remark we want to emphasize that the method of writing the
Hamiltonian in terms of the Casimirs can be applied to systems with any number
of particles (at the price, increasing their number, of a growing complexity
in the induced irreducible representations) and moreover the analysis
may not be restricted to singlet subspaces.
Unfortunately it is not clear how to define a method for interacting particle
not in the bifundamental representation. 

%
\end{document}